\documentclass[preprint,12pt]{elsarticle}
\usepackage{graphicx} 
\usepackage{amssymb}
\usepackage{amsmath}
\usepackage{amsfonts}
\usepackage{slashed}
\usepackage[dvipsnames]{xcolor}
\usepackage{braket,bm}
\usepackage{multirow}
\usepackage{enumitem}
\usepackage[vcentermath]{youngtab}
\def\lamb#1#2{$^{#1}_{\Lambda}${#2}}

\def\lamblamb#1#2{$^{~~#1}_{\Lambda\Lambda}${#2}}

\newcommand{\bs}[1]{\boldsymbol{#1}} 
\newcommand{\nopieft}{\mbox{$\slashed{\pi}$EFT}}

\newcommand{\be}{\begin{equation}} 
\newcommand{\ee}{\end{equation}}
\newcommand{\rvec}{{\bs{r}}}
\newcommand{\Rvec}{{\bs{R}}}

\newcommand{\kvec}{{\bs{k}}} 
 
\newcommand{\lamvec}{\ensuremath{\boldsymbol{\lambda}}} 
\newcommand{\rhovec}{\ensuremath{\boldsymbol{\rho}}}

\journal{Physics Letters B} 

\begin{document}

\begin{frontmatter}

\title{Hypernuclear constraints on the existence and lifetime of a deeply 
bound $H$ dibaryon} 
\author{Avraham Gal\corref{cor}~} 
\address{Racah Institute of Physics, The Hebrew University, 
Jerusalem 9190401, Israel} 
\cortext[cor]{Avraham Gal,~~avragal@savion.huji.ac.il}
\date{\today}

\begin{abstract} 
We study to what extent the unique observation of $\Lambda\Lambda$ hypernuclei 
by their weak decay into known $\Lambda$ hypernuclei, with lifetimes of 
order 10$^{-10}$~s, rules out the existence of a deeply bound doubly-strange 
(${\cal S}$=$-$2) $H$ dibaryon. Treating \lamblamb{6}{He} (the Nagara emulsion 
event) in a realistic $\Lambda-\Lambda-{^4}$He three-body model, we find that 
the ${_{\Lambda\Lambda}^{~~6}}{\rm He}\to H + {^4{\rm He}}$ strong-interaction 
lifetime increases beyond 10$^{-10}$~s for $m_H < m_{\Lambda}+m_n$, about 176 
MeV below the $\Lambda\Lambda$ threshold, so that such a deeply bound $H$ is 
not in conflict with hypernuclear data. Constrained by $\Lambda$ hypernuclear 
$\Delta{\cal S}$=1 nonmesonic weak-interaction decay rates, we follow EFT 
methods to evaluate the $\Delta{\cal S}$=2 $H\to nn$ weak-decay lifetime of 
$H$ in the mass range $2m_n \lesssim m_H < m_{\Lambda}+m_n$. The resulting $H$ 
lifetime is of order 10$^5$~s, many orders of magnitude shorter than required 
to qualify for a dark-matter candidate. 
\end{abstract} 

\end{frontmatter} 

\section{Introduction} 
\label{sec:intro} 

The deuteron, with mass only 2.2~MeV below the sum of masses of its proton 
and neutron constituents, is the only particle-stable six-quark (hexaquark) 
dibaryon known so far. Here stabilty is regarded with respect to the lifetime 
of the proton, many orders of magnitude longer than the lifetime of the 
Universe (13.8 billion years~\cite{AA20}). Extending the very-light $ud$ 
quark sector by the light strange quark $s$, Lattice-QCD (LQCD) calculations 
suggest two strong-interaction stable hexaquarks. Both are $J^{\pi}$=0$^+$ 
near-threshold $s$-wave dibaryons with zero spin and isospin: 
(i) a maximally strange ${\cal S}$=$-$6 $ssssss$ hexaquark classified as 
$\Omega\Omega$ dibaryon member of the SU(3) flavor ${\bf 28}_f$ multiplet, 
and (ii) a strangeness ${\cal S}$=$-$2 $uuddss$ hexaquark, a ${\bf 1}_f$ 
$H$ dibaryon, which is the subject of the present study. Whereas the LQCD 
calculation of $\Omega\Omega$ reached $m_{\pi}$ values close to the physical 
pion mass~\cite{LQCD18}, $H$ dibaryon LQCD calculations have been limited 
to values of $m_{\pi}\sim 400$~MeV and higher (NPLQCD~\cite{LQCD11a}, 
HALQCD~\cite{LQCD11b}) while following SU(3)$_f$ symmetry, where 
\begin{equation} 
H = -\sqrt{\frac{1}{8}}\Lambda\Lambda +\sqrt{\frac{3}{8}}\Sigma\Sigma 
+\sqrt{\frac{4}{8}}N\Xi. 
\label{eq:H} 
\end{equation} 
A very recent calculation of this type~\cite{LQCD21} finds the $H$ dibaryon 
bound just by 4.6$\pm$1.3~MeV with respect to the $\Lambda\Lambda$ threshold. 
However, chiral extrapolation to physical quark masses values and thereby also 
to $m_{\pi}\approx 0$~\cite{LQCD11c} suggests that the $H$ dibaryon becomes 
{\it unbound} by 13$\pm$14~MeV. Thus, a slightly bound $\bf{1}_f$ $H$ dibaryon 
is likely to become unbound with respect to the $\Lambda\Lambda$ threshold in 
the SU(3)$_f$-broken physical world, lying possibly a few MeV below the $N\Xi$ 
threshold~\cite{LQCD12,EFT12}. 

The $H$ dibaryon was predicted in 1977 by Jaffe~\cite{Jaffe77} to 
lie about 80~MeV below the 2$m_{\Lambda}$=2231~MeV $\Lambda\Lambda$ 
threshold. Dedicated experimental searches, beginning as soon as 1978 with 
a $pp\to K^+K^-X$ reaction~\cite{AGS78} at BNL, have failed to observe 
a ${\cal S}$=$-$2 dibaryon signal over a wide range of dibaryon masses below 
2$m_{\Lambda}$~\cite{Belle13,ALICE16,BABAR19}, notably BABAR's recent search 
at SLAC looking for a $\Upsilon(2S,3S)\to{\bar\Lambda}{\bar\Lambda}X$ 
decay~\cite{BABAR19}. Furthermore, a simple argument questioning the existence 
of a strong-interaction stable ${\cal S}$=$-$2 dibaryon was put forward by 
several authors, notably Dalitz et al.~\cite{Dalitz89}. It relates to the 
few established $\Lambda\Lambda$ hypernuclei~\cite{GHM16,HN18}, foremost 
the lightest known one \lamblamb{6}{He} (the Nagara emulsion event) where 
a $\Lambda\Lambda$ pair is bound to $^4$He by 6.91$\pm$0.17~MeV~\cite{Ahn13} 
exceeding twice the separation energy of a single $\Lambda$ in \lamb{5}{He} 
by merely $\Delta B_{\Lambda\Lambda}(^{~~6}_{\Lambda\Lambda}$He)=0.67$\pm
$0.17~MeV. If $H$ existed deeper than about 7~MeV below the $\Lambda\Lambda$ 
threshold, \lamblamb{6}{He} could decay {\it strongly}, 
\begin{equation} 
{_{\Lambda\Lambda}^{~~6}}{\rm He} \to {^4{\rm He}} + H,  
\label{eq:LL6He} 
\end{equation} 
considerably faster than the $\Delta{\cal S}$=1 {\it weak-interaction} decay 
by which it has been observed and uniquely identified~\cite{Ahn13}: 
\begin{equation} 
{_{\Lambda\Lambda}^{~~6}}{\rm He} \to {_{\Lambda}^5}{\rm He}+p+\pi^-. 
\label{eq:L5He}
\end{equation} 
Further arguments questioning an $H$ bound by less than about 7~MeV were put 
forward by Gal~\cite{Gal13}. 

Arguments of this kind, questioning the existence of a strong-interaction 
stable $H$ dibaryon, were challenged 20 years ago by Farrar~\cite{Farrar03} 
who suggested that the $H$ dibaryon may be a long-lived compact object with 
as small radius as 0.2~fm and as small mass as 1.5$\pm$0.2~GeV, in which case 
it becomes absolutely stable, without disrupting the observed stability of 
nuclei. Once sufficiently abundant, relic $H$ dibaryons would qualify as 
a cold Dark Matter (DM) candidate. Farrar's present estimate for the mass 
$m_H$ of such a compact dibaryon, often termed Sexaquark in these works, 
is between 1850 and 2050~MeV~\cite{FW23}. Here, $m_H\lesssim 1850$~MeV is 
disfavored by the stability of oxygen~\cite{SK15}, whereas the mass value 
2050~MeV stems from the threshold value $(m_n+m_{\Lambda})=2055$~MeV, 
above which the $\Delta \cal{S}$=1 strangeness changing weak decay 
$H\to n \Lambda$ would make $H$ definitely short lived with respect 
to a lifetime of cosmological origin expected for a DM candidate. 

Following Farrar's conjecture of a deeply-bound compact ${\cal S}$=$-$2 
dibaryon, we present a realistic calculation of \lamblamb{6}{He} lifetime 
owing to the two-body strong-interaction decay reaction Eq.~(\ref{eq:LL6He}). 
Treating \lamblamb{6}{He} in a $\Lambda-\Lambda-{^4}$He three-body model, 
it is found that the ${_{\Lambda\Lambda}^{~~6}}{\rm He}\to H + {^4{\rm He}}$ 
strong-interaction lifetime is correlated strongly with $m_H$, increasing 
upon decreasing $m_H$ such that it exceeds the hypernuclear $\Delta\cal{S}$=1 
weak-decay lifetime scale of order 10$^{-10}$~s for $m_H < (m_{\Lambda}+m_n)$. 
Therefore, hypernuclear physics by itself does not rule out an $H$-like 
dibaryon in this mass range. 

Constrained by $\Lambda$ hypernuclear $\Delta{\cal S}$=1 nonmesonic 
weak-interaction decay rates within leading-order (LO) effective field theory 
(EFT) approach, we present a realistic calculation of the $\Delta\cal{S}$=2 
weak decay $H\to nn$ for $H$ mass satisfying $2m_n \lesssim m_H < (m_n+m_{
\Lambda})$. The resulting $H$ lifetimes are of order $10^5$~s, in rough 
agreement with Donoghue, Golowich, Holstein \cite{DGH86} who followed 
a completely different high-energy physics methodology. Our calculated $H$ 
lifetimes are 10 orders of magnitude shorter than the order of 10$^8$ yr 
reached in the 2004 Farrar-Zaharijas (FZ) calculation~\cite{FZ04}. 
Hence, a deeply bound $H$ dibaryon would be far from qualifying for a DM 
candidate.

\section{$H$ dibaryon wavefunction} 
\label{sec:Hwf} 

\subsection{Spatial part} 
\label{subsec:space} 

Here we follow the simple ansatz for the $H$ dibaryon six-quark (6$q$) fully 
symmetric spatial wavefunction $\Psi_H$ given by FZ~\cite{FZ04}: 
\begin{equation} 
\Psi_H = N_6\exp \left(-\frac{\nu}{6}\sum_{i<j}^{6}
({\rvec_i}-{\rvec_j})^2 \right)\,, 
\label{eq:PsiH} 
\end{equation} 
where $N_6$ is a normalization constant and $\nu$ is related to the $H$ `size' 
as detailed below. To transform this 6$q$ $\Psi_H$ to a two-baryon form where 
each baryon $B_a$ and $B_b$ is described as a 3$q$ cluster, we define relative 
coordinates $\rhovec,\,\lamvec$ and center-of-mass (c.m.) coordinates 
$\Rvec$: 
\begin{equation} 
B_a:\,\,\,\rhovec_a=\rvec_2 - \rvec_1,\,\,\,\,\,\, 
\lamvec_a=\rvec_3 - \frac{1}{2}(\rvec_1 + \rvec_2),\,\,\,\,\,\, 
\Rvec_a = \frac{1}{3}(\rvec_1 + \rvec_2 + \rvec_3),   
\label{eq:a} 
\end{equation} 
\begin{equation} 
B_b:\,\,\,\rhovec_b=\rvec_5 - \rvec_4,\,\,\,\,\,\,
\lamvec_b=\rvec_6 - \frac{1}{2}(\rvec_4 + \rvec_5),\,\,\,\,\,\, 
\Rvec_b = \frac{1}{3}(\rvec_4 + \rvec_5 + \rvec_6), 
\label{eq:b} 
\end{equation} 
plus a total cm coordinate $\Rvec = \frac{1}{2}(\Rvec_a + \Rvec_b)
=\frac{1}{6}\sum_i^6{\rvec_i}$. Using these $\rhovec$ and $\lamvec$ intrinsic 
quark coordinates, plus a relative coordinate $\rvec = (\Rvec_b-\Rvec_a)$ 
between the baryonic 3$q$ clusters $B_a$ and $B_b$, Eq.~(\ref{eq:PsiH}) 
assumes the form 
\begin{equation} 
\Psi_H=\psi_{B_a}(\rhovec_a,\lamvec_a)\times\psi_{B_b}(\rhovec_b,\lamvec_b) 
\times\psi_{B_aB_b}(\rvec), 
\label{eq:ab} 
\end{equation}
where 
\begin{equation} 
\psi_{B_j}=\left(\frac{4{\nu}^2}{3{\pi}^2}\right)^{\frac{3}{4}}
\exp (-\frac{\nu}{2}\rho_j^2-\frac{2\nu}{3}\lambda_j^2) 
\label{eq:Bj} 
\end{equation} 
provides normalized 3$q$ baryonic spatial wavefunction for baryon $B_j$, 
$j=a,b$, and 
\begin{equation} 
\psi_{B_aB_b}=\left(\frac{3\nu}{\pi}
\right)^{\frac{3}{4}}\exp(-\frac{3\nu}{2}r^2) 
\label{eq:Bab} 
\end{equation} 
provides a normalized spatial wavefunction in the relative coordinate 
$\vec r$ of the dibaryon $B_aB_b$. Note that all three components of $H$ in 
Eq.~(\ref{eq:ab}) share the {\it same} root-mean-square (r.m.s.) radius value: 
\begin{equation} 
<r^2_{B_a}>=<r^2_{B_b}>=<r^2_{B_aB_b}>=\frac{9}{8\nu}, 
\label{eq:sizeBj} 
\end{equation} 
while for $\Psi_H$ it is given by 
\begin{equation} 
<r^2_H> = \frac{5}{8\nu} 
\label{eq:sizeH} 
\end{equation} 
(correcting an error: $<r^2_H>=\nu^{-1}$ in Ref.~\cite{FZ04}), so that the 
radial extent of $H$ is about 75\% of the radial extent of each of its three 
components $B_a,B_b,B_aB_b$. We also note that a 3$q$ baryon wavefunction, 
similar to $\Psi_H$ for 6$q$, 
\begin{equation} 
\Psi_B = N_3\exp \left(-\frac{\nu}{6}\sum_{i<j}^{3}
({\rvec_i}-{\rvec_j})^2 \right), 
\label{eq:PsiB} 
\end{equation} 
implies a r.m.s. radius squared of 
\begin{equation} 
<r_B^2> = \frac{1}{\nu}, 
\label{eq:sizeB} 
\end{equation} 
slightly smaller according to Eq.~(\ref{eq:sizeBj}) than when embedded within 
the $H$ dibaryon. 

\begin{table}[!t] 
\begin{center} 
\caption{$\sqrt{<r^2_{\Lambda\Lambda}>}$ (fm units) vs. $B_{\Lambda\Lambda}$ 
(MeV units) for a short-range Gaussian $\Lambda\Lambda$ potential, 
Eq.~(\ref{eq:gaussian}) with $\lambda = 4$~fm$^{-1}$. I'm indebted to Martin 
Sch{\"a}fer for providing me with this table.} 
\begin{tabular}{ccccccccc} 
\hline 
$B_{\Lambda\Lambda}$ & 5 & 20 & 50 & 100 & 200 & 300 & 400 & 1000 \\ 
\hline 
$\sqrt{<r^2_{\Lambda\Lambda}>}$ & 2.134 & 1.206 & 0.854 & 0.689 & 
0.560 & 0.501 & 0.463 & 0.366 \\ 
\hline
\end{tabular} 
\label{tab:LLsize} 
\end{center} 
\end{table} 

Although derived for a specific spatially {\it symmetric} wavefunction, 
Eq.~(\ref{eq:PsiH}), the relationships noted above between various $<r^2>$ 
values hold for {\it any} spatially symmetric form chosen for $H$. 
Establishing physically one such `size' value determines necessarily 
all other `size' values. However, the choice of a specific `size' value is 
constrained by the choice of $H$ binding energy value, as demonstrated in 
Table~\ref{tab:LLsize} for $B_a=B_b=\Lambda$. In this particular case, the 
Schroedinger equation in the relative coordinate $\rvec_{\Lambda\Lambda}$ 
was solved for assumed binding energy values $B_{\Lambda\Lambda}$, 
using attractive Gaussian $\Lambda\Lambda$ potential of the form 
$C_0^{(\lambda)}\delta_\lambda(\rvec)$, where $C_0^{(\lambda)}$ 
is a strength parameter fitted to given values of $B_{\Lambda\Lambda}$ and 
\begin{equation}
\delta_\lambda(\rvec)=\left(\frac{\lambda}{2\sqrt{\pi}}\right)^3\,
\exp \left(-{\frac{\lambda^2}{4}}\rvec^2\right)   
\label{eq:gaussian} 
\end{equation} 
is a zero-range Dirac $\delta^{(3)}(\rvec)$ function in the limit $\lambda
\to\infty$, smeared over distance of $\sqrt{<r^2>_{\lambda}}=\sqrt{6}/
\lambda$ (0.612~fm for $\lambda = 4$~fm$^{-1}$ chosen here). As expected, 
once $B_{\Lambda\Lambda}$ increases beyond a nuclear physics scale of roughly 
20 MeV or so, $\sqrt{<r^2_{\Lambda\Lambda}>}$ decreases below 1~fm down to 
$\approx 0.5$~fm. The corresponding $H$ r.m.s. radius values are even smaller: 
$\sqrt{<r^2_H>}/\sqrt{<r^2_{\Lambda\Lambda}>}=(\sqrt{5}/3)\approx 0.745$ 
according to Eqs.~(\ref{eq:sizeBj},\ref{eq:sizeH}). Taking a shorter-range 
Gaussian potential, say with $\lambda = 5$~fm$^{-1}$, has a relatively small 
effect on $\sqrt{<r^2_{\Lambda\Lambda}>}$ which decreases between 3\% to 12\% 
as $B_{\Lambda\Lambda}$ increases from 5 to 1000~MeV. In passing we comment 
that the constraint imposed on $\sqrt{<r^2_H>}$ by assuming a definite value 
of $B_{\Lambda\Lambda}$, or vice versa, was overlooked in Ref.~\cite{FZ04}. 

In the calculations reported below we use the fully symmetric spatial $H$ 
wavefunction (\ref{eq:PsiH}) or equivalently (\ref{eq:ab}) by choosing 
$\nu=(9/8){<r^2_{\Lambda\Lambda}>}^{-1}$, see Eq.~(\ref{eq:sizeBj}), where 
$<r^2_{\Lambda\Lambda}>$ is obtained by solving the Schroedinger equation 
with attractive $\Lambda\Lambda$ potential shape, Eq.~(\ref{eq:gaussian}) for 
$\lambda = 4$~fm$^{-1}$, and variable strength determined by assuming given 
values of $B_{\Lambda\Lambda}$, as listed in Table~\ref{tab:LLsize}.

\subsection{Spin-Flavor-Color part} 
\label{subsec:SFC}

To complete the discussion of the $H$ dibaryon wavefunction we note that the 
fully symmetric spatial 6$q$ wavefunction $\Psi_H$, Eq.~(\ref{eq:PsiH}), needs 
to be supplemented by a singlet $\textbf{1}_S$ total spin $S$=0 component, 
represented by a 6$q$ SU(2) Young tableau 
\begin{equation} 
{\yng(3,3)}_{~S}\,, 
\label{eq:S} 
\end{equation} 
and by singlet $\textbf{1}_F$ total flavor ($F$) and $\textbf{1}_C$ total 
color ($C$) components, each represented by its own 6$q$ SU(3) Young tableau 
\begin{equation} 
{\yng(2,2,2)}_{~F~{\rm or}~C}\,.  
\label{eq:FC} 
\end{equation} 
Each of these $S$, $F$ and $C$ tableaux accommodates five components. In 
spin space, only one component corresponds to $S_a(uds)=S_b(uds)=\frac{1}{2}$ 
{\textit and} $S_a(ud)=S_b(ud)=0$ implied by a $\Lambda\Lambda$ dibaryon 
component, and in flavor space, again, only one component corresponds to 
$\textbf{8}_a(uds)$ and $\textbf{8}_b(uds)$ with isospin $I_a(ud)=I_b(ud)=0$ 
for a $\Lambda\Lambda$ dibaryon component. In color space, too, only one 
component corresponds to colorless $\textbf{1}_a(uds)$ and $\textbf{1}_b(uds)$ 
baryons. Hence, up to a phase, we assign in each of these three spaces 
a coefficient of fractional parentage $\sqrt{1/5}$ to $\Psi_H$ of 
Eq.~(\ref{eq:ab}). Finally, having chosen the $\Lambda\Lambda$ component over 
the $\Sigma\Sigma$ and $N\Xi$ components of $H$, see Eq.~(\ref{eq:H}), 
involves a Clebsch-Gordan coefficient of magnitude $\sqrt{1/8}$ which together 
with the former coefficients amounts to supplementing the spatially symmetric 
$\Lambda\Lambda$ wavefunction $\psi_{\Lambda\Lambda}$, Eq.~(\ref{eq:ab}) 
for $B_a=B_b=\Lambda$, by a flavor-color-spin factor $\sqrt{1/1000}$: 
\begin{equation} 
{\tilde{\psi}}_{\Lambda\Lambda}=\sqrt{1/1000}\times\psi_{\Lambda\Lambda}.  
\label{eq:psitilde} 
\end{equation} 
Representing the $H$ dibaryon spatially by the fairly small-size 
${\tilde{\psi}}_{\Lambda\Lambda}$ wavefunction rather than by the normal-size 
normalized $\psi_{\Lambda\Lambda}$ means that its initial- and final-state 
interactions with `normal' baryonic matter are negligible, in agreement with 
arguments reviewed in Ref.~\cite{FW23}. Accordingly, no final-state 
interaction between $H$ and ${^4}$He is introduced in the strong-interaction 
decay \lamblamb{6}{He} $\to H + {^4}$He studied in Sect.~\ref{sec:decayrate} 
below.

\section{\lamblamb{6}{He} wavefunction} 
\label{sec:LL6Hewf} 

\subsection{Three-body approximation} 
\label{subsec:approx} 

Given the tight binding of $^4$He, we treat the six-body \lamblamb{6}{He} 
as a three-body $\Lambda\Lambda\alpha$ system with spatial coordinates 
${\rvec}_{\alpha}$, ${\rvec}_{\Lambda_1}$, ${\rvec}_{\Lambda_2}$. 
Starting from the two relative $\Lambda\alpha$ vector coordinates 
${\rvec}_{\Lambda_1\alpha} = {\rvec}_{\Lambda_1}-{\rvec}_{\alpha}$ 
and ${\rvec}_{\Lambda_2\alpha} = {\rvec}_{\Lambda_2}-{\rvec}_{\alpha}$, 
we transform to their relative and c.m. coordinates 
\begin{equation} 
{\rvec}_{\Lambda\Lambda}={\rvec}_{\Lambda_2\alpha}-{\rvec}_{\Lambda_1\alpha}, 
\,\,\,\,\,\,{\Rvec}_{\Lambda\Lambda}=\frac{1}{2}
({\rvec}_{\Lambda_1\alpha}+{\rvec}_{\Lambda_2\alpha}). 
\label{eq:rel} 
\end{equation} 
A reasonable simple approximation of the Pionless-EFT ({\nopieft}) 
\lamblamb{6}{He} wavefunction calculated in Ref.~\cite{CSBGM19} is 
then to use a factorized ansatz: 
\begin{equation} 
\Phi_{{^{~~6}_{\Lambda\Lambda}}{\rm He}}=\phi_{\Lambda\Lambda}(r_{\Lambda
\Lambda})\,\Phi_{\Lambda\Lambda}(R_{\Lambda\Lambda})\,\phi_{\alpha}, 
\label{eq:PhiLL6He} 
\end{equation} 
where the wavefunctions $\phi_{\Lambda\Lambda}$ and $\Phi_{\Lambda\Lambda}$ 
are chosen as Gaussians constrained by requiring that $\phi_{\Lambda\Lambda}$ 
reproduces the r.m.s. radius of the coordinate ${\rvec}_{\Lambda\Lambda}$ 
in the 6-body \lamblamb{6}{He} {\nopieft} calculation~\cite{CSBGM19} as 
discussed below. Note that the r.m.s. radius value of the c.m. Gaussian
$\Phi_{\Lambda\Lambda}$ is half that of the Gaussian $\phi_{\Lambda\Lambda}$. 
Finally, the $^4$He core wavefunction $\phi_{\alpha}$ within \lamblamb{6}{He} 
is approximated by a free-space $^4$He wavefunction identical with that for 
$^4$He in the ${^{~~6}_{\Lambda\Lambda}}{\rm He}\to H+{^4{\rm He}}$ 
strong-interaction decay. 

Studying the {\nopieft} \lamb{5}{He} five-body calculation~\cite{CBG18} we 
note that $B_{\Lambda}^{\rm exp}({_{\Lambda}^5{\rm He}})$ is nearly reproduced 
by choosing Eq.~(\ref{eq:gaussian}) for $\Lambda N$ contact terms, with cutoff 
values $\lambda=1.25$~fm$^{-1}$ or $\lambda=1.50$~fm$^{-1}$ for $\Lambda N$ 
scattering length versions Alexander(B) and $\chi$EFT(NLO19). Going over 
to the {\nopieft} \lamblamb{6}{He} six-body calculation~\cite{CSBGM19}, 
the $\Lambda\Lambda$ r.m.s. distance computed for these cutoff values is  
$\sqrt{<r^2_{\Lambda\Lambda}>}=3.65\pm 0.10$~fm, which we adopt for the r.m.s. 
radius of the Gaussian $\phi_{\Lambda\Lambda}$ in Eq.~(\ref{eq:PhiLL6He}). 
Note that $\phi_{\Lambda\Lambda}$ appears as a bound-state wavefunction in 
spite of the $\Lambda\Lambda$ interaction being much too weak to form a bound 
state; it is the $^4$He nuclear core that stabilizes the two $\Lambda$s in 
\lamblamb{6}{He}. 
We also note that since this value refers to a weakly `bound' $\Lambda\Lambda$ 
pair in \lamblamb{6}{He}, it is considerably larger than $\sqrt{<r_{\Lambda
\Lambda}^2>}$ values listed in Table~\ref{tab:LLsize} for a tightly bound $H$ 
dibaryon.

\subsection{Short-range behavior} 
\label{subsec:SRC} 

Eq.~(\ref{eq:PhiLL6He}) provides a simple wavefunction for two loosely bound 
$\Lambda$ hyperons held together by ${^4}$He, disregarding the short-range 
repulsive component of the $\Lambda\Lambda$ interaction which is manifest in 
LQCD calculations~\cite{Inoue19}. To account for the short-range repulsion 
effect on the \lamblamb{6}{He} $\to H+{^4}$He decay rate, we modify 
$\phi_{\Lambda\Lambda}$ in Eq.~(\ref{eq:PhiLL6He}) by introducing 
a short-range correlation (SRC) factor $[1-j_0({\kappa}r)]$, where $j_0$ is 
a spherical Bessel function of order zero: 
\begin{equation} 
{\tilde{\phi}}_{\Lambda\Lambda}(r_{\Lambda\Lambda})=
\left(1-j_0({\kappa}r_{\Lambda\Lambda})\right)\,\phi_{\Lambda\Lambda}
(r_{\Lambda\Lambda}). 
\label{eq:phiLL} 
\end{equation} 
Choosing $\kappa = 2.534$~fm$^{-1}$, corresponding to 500 MeV/c in momentum 
space, nearly reproduces the $\Lambda\Lambda$ $G$-matrix calculation in 
Ref.~\cite{MPR18} (Fig.~2 there and related text). We therefore replace 
$\Phi_{{^{~~6}_{\Lambda\Lambda}}{\rm He}}$, Eq.~(\ref{eq:PhiLL6He}), by 
\begin{equation} 
\Psi_i={\tilde{\phi}}_{\Lambda\Lambda}(r_{\Lambda\Lambda})\,
\Phi_{\Lambda\Lambda}(R_{\Lambda\Lambda})\,\phi_{\alpha} 
\label{eq:Psi_i} 
\end{equation} 
for use as initial \lamblamb{6}{He} wavefunction in the \lamblamb{6}{He} 
$\to H + {^4}$He decay rate calculation reported below.

\section{\lamblamb{6}{He} $\to H + {^4}$He decay rate} 
\label{sec:decayrate} 

We assume that the strong-interaction decay \lamblamb{6}{He} $\to H+{^4}$He 
of a loosely `bound' $\Lambda\Lambda$ pair in \lamblamb{6}{He} into a 
$\Lambda\Lambda$ pair constituent of a tightly bound $\textbf{1}_F$ $H$ 
dibaryon, flying off $^4$He with momentum ${\kvec}_H$ in their c.m. system, 
is triggered by the $\Lambda\Lambda$ strong interaction $V_{\Lambda\Lambda}$ 
extracted near threshold. The spatial dependence of the decay matrix element 
is given by $<\Psi_f|V_{\Lambda\Lambda}|\Psi_i>$, where $\Psi_i$ stands for 
the initial \lamblamb{6}{He} wavefunction, Eq.~(\ref{eq:Psi_i}), and 
\begin{equation} 
\Psi_f=\tilde\psi_{\Lambda\Lambda}(r_{\Lambda\Lambda})
\exp{(i{\kvec}_H\cdot{\Rvec}_H)}\phi_{\alpha},\;\;\;\;\;
{\tilde{\psi}}_{\Lambda\Lambda}=\sqrt{1/1000}\times\psi_{\Lambda\Lambda}, 
\label{eq:Psi_f} 
\end{equation} 
where $\psi_{\Lambda\Lambda}$, Eq.~(\ref{eq:Bab}), is renormalized by the 
flavor-color-spin factor $\sqrt{1/1000}$, see Eq.~(\ref{eq:psitilde}), thereby 
accounting for the elimination of $\Sigma\Sigma$ and $N\Xi$ components of $H$. 
Note that in agreement with the overall attraction of the $BB$ interaction in 
the $\textbf{1}_F$ channel~\cite{Inoue19} no SRC factor was introduced in 
Eq.~(\ref{eq:Psi_f}) for $\Psi_f$. We note that the calculations reported 
below disregard the slight difference between the inner 3$q$ structure of 
each $\Lambda$ hyperon in the $H$ dibaryon to that in \lamblamb{6}{He}. 
For a 3$q$ baryon size of about 0.5~fm~\cite{KW24}, this neglect is well 
justified in the range of $B_{\Lambda\Lambda}$ values considered here. 

The \lamblamb{6}{He} $\to H+{^4}$He decay rate, or equivalently 
the corresponding strong-interaction width of \lamblamb{6}{He}, is given 
by~\cite{Baym69}:  
\begin{equation} 
\Gamma({_{\Lambda\Lambda}^{~6}{\rm He}}\to H + {{^4}{\rm He}}) = 
\frac{\mu_{H\alpha}\,k_H}{(2\pi\hbar c)^2} \int{
|<\Psi_f|V_{\Lambda\Lambda}|\Psi_i>|^2\,{\rm d}{\hat\kvec}_H}, 
\label{eq:LL6Hewidth} 
\end{equation}
where $\mu_{H\alpha}$ is the $H-{^{4}{\rm He}}$ reduced mass and 
$<\Psi_f|V_{\Lambda\Lambda}|\Psi_i>$ is a product of two factors, as follows. 

One factor is the $\Lambda\Lambda$ interaction matrix element $<\tilde{\psi}_{
\Lambda\Lambda}|V_{\Lambda\Lambda}|\tilde{\phi}_{\Lambda\Lambda}>$ in the 
$r_{\Lambda\Lambda}$ relative distance with $V_{\Lambda\Lambda}$ connecting 
$\tilde{\phi}_{\Lambda\Lambda}$, Eq.~(\ref{eq:phiLL}), for the initial 
\lamblamb{6}{He} wavefunction component to the final $H$ dibaryon 
renormalized wavefunction component ${\tilde{\psi}}_{\Lambda\Lambda}$, 
Eq.~(\ref{eq:Psi_f}). A normalized Gaussian $\delta_{\lambda=4}(\rvec)$, 
see Eq.~(\ref{eq:gaussian}), was used for $V_{\Lambda\Lambda}$ with 
strength parementer $C_0^{(\lambda=4)}=-152$~MeV$\cdot$fm$^3$ fitted in 
Ref.~\cite{CSBGM19} to the HAL-QCD scattering length $a_{\Lambda\Lambda}=-0.8
$~fm~\cite{Sasaki20}. The calculated matrix element $<V_{\Lambda\Lambda}>$ 
depends weakly on the chosen value of $\lambda$ within $\delta\lambda=\pm 1
$~fm$^{-1}$. As for dependence on SRC, Eq.~(\ref{eq:phiLL}), it 
introduces a multiplicative factor 
\begin{equation} 
1-\exp{(-{\kappa}^2/4\chi)} 
\label{eq:SR} 
\end{equation} 
to $<V_{\Lambda\Lambda}>$, with $\chi=\frac{1}{2}(\nu_i+\nu_f)+\frac{1}{4}
\lambda^2$ for wavefunctions of Gaussian shape $\exp(-\nu r^2/2)$. This factor 
varies from 0.25 to 0.19 as $B_{\Lambda\Lambda}$ is increased from 100 to 
400~MeV. Altogether, the matrix element $<\tilde{\psi}_{\Lambda\Lambda}|
V_{\Lambda\Lambda}|{\tilde\phi}_{\Lambda\Lambda}>$ varies slightly from 
$-$59 to $-$53~keV in the same $B_{\Lambda\Lambda}$ range. 

\begin{table}[!h] 
\begin{center} 
\caption{\lamblamb{6}{He} $\to H + {^4}$He decay rate $\Gamma$, 
Eq.~(\ref{eq:LL6Hewidth}), and decay time $\hbar/\Gamma$ for some 
representative values of $H$ binding energy $B_{\Lambda\Lambda}$ and their 
associated $k_H$ and $I(k_H;a_{\Phi})$ values for $a_{\Phi}=1.49$~fm, 
see text. $B_{\Lambda\Lambda}=176$~MeV corresponds to $m_H=m_{\Lambda}+m_n$.} 
\begin{tabular}{ccccc} 
\hline 
$B_{\Lambda\Lambda}$ (MeV) & $k_H$ (fm$^{-1}$) & $I(k_H;a_{\Phi})$ (fm$^3$) 
& $\Gamma$ (eV) & $\tau=\hbar/\Gamma$ (s) \\ 
\hline 
100 & 2.547 & $1.002\cdot 10^{-3}$ & $0.782\cdot 10^{-2}$ & 
$0.841\cdot 10^{-13}$ \\ 
200 & 3.612 & $4.742\cdot 10^{-10}$ & $0.501\cdot 10^{-8}$ & 
$1.315\cdot 10^{-7}$ \\ 
300 & 4.377 & $5.996\cdot 10^{-16}$ & $0.679\cdot 10^{-14}$ & 
$0.970\cdot 10^{-1}$ \\ 
400 & 4.980 & $2.157\cdot 10^{-21}$ & $2.436\cdot 10^{-20}$ & 
$2.703\cdot 10^4$ \\
\hline
\hline 
176 & 3.393 & $1.521\cdot 10^{-8}$ & $1.550\cdot 10^{-7}$ & 
$4.245\cdot 10^{-9}$ \\ 
\hline
\end{tabular} 
\label{tab:I_R} 
\end{center} 
\end{table} 

The other factor in $<\Psi_f|V_{\Lambda\Lambda}|\Psi_i>$ is an overlap matrix 
element between the initial $\Lambda\Lambda - \alpha$ Gaussian wavefunction 
$\Phi_{\Lambda\Lambda}(R_{\Lambda\Lambda})$ in \lamblamb{6}{He} and the final 
outgoing $H - \alpha$ plane-wave $\exp{(i{\kvec}_H \cdot{\Rvec}_H)}$. Note 
that ${\Rvec}_H$ has nothing to do with the relatively small size of $H$. 
In the following we identify ${\Rvec}_H$ with the corresponding argument 
${\Rvec}_{\Lambda\Lambda}$ in Eq.~(\ref{eq:PhiLL6He}), both defined relative 
to $^4$He and denoted below simply by ${\Rvec}$. The square of this overlap 
matrix element times $4\pi$ from d${\hat\kvec}_H$ in Eq.~(\ref{eq:LL6Hewidth}) 
is given by 
\begin{equation} 
I(k_H;a_{\Phi})\equiv 4\pi{\left |\,\int{\exp(i{\kvec}_H\cdot{\Rvec})
\Phi_{\Lambda\Lambda}(R){\rm d}^3\Rvec}\,\right|}^2 = 
32{\pi}^{5/2}a_{\Phi}^3\exp(-a_{\Phi}^2k_H^2), 
\label{eq:I_R} 
\end{equation} 
where $a_{\Phi}=\sqrt{2<R_{\Lambda\Lambda}^2>/3}=1.49\pm 0.04$~fm. As shown 
in Table~\ref{tab:I_R}, $I(k_H;a_{\Phi})$ varies strongly with $k_H$ over 
18 decades as $B_{\Lambda\Lambda}$ is varied from 100~MeV (76~MeV above 
$m_{\Lambda}+m_n$) to 400~MeV (47~MeV below 2$m_n$). This is caused by the 
increased oscillations of $\exp(i{\kvec}_H\cdot{\Rvec})$ vs. the smoothly 
varying $\Phi_{\Lambda\Lambda}(R)$ in Eq.~(\ref{eq:I_R}). And on top of that, 
the $\pm 0.04$~fm uncertainty of $a_{\Phi}$ makes $I(k_H;a_{\Phi})$ uncertain 
by a factor of 4 larger/smaller values than for the mean value $a_{\Phi}=1.49
$~fm at $B_{\Lambda\Lambda}=176$~MeV corresponding to the $m_{\Lambda}+m_n$ 
threshold, increasing to a factor about 20 at $B_{\Lambda\Lambda}=400$~MeV. 

The final values of \lamblamb{6}{He} $\to H + {^4}$He decay rate $\Gamma$ 
listed in Table~\ref{tab:I_R} account for the two factors in $<\Psi_f|V_{
\Lambda\Lambda}|\Psi_i>$ considered above. Notably, $\Gamma$ decreases over 
17 decades, reflecting the strong variation of $I(k_H;a_{\Phi})$ with 
$B_{\Lambda\Lambda}$, and the \lamblamb{6}{He} strong-interaction lifetime 
$\tau=\hbar/\Gamma$ increases as strongly over this range of $B_{\Lambda
\Lambda}$ values. In particular, for $H$ binding energy $B_{\Lambda\Lambda}= 
176$~MeV ($m_{\Lambda}+m_n$ threshold), and given the $a_{\Phi}$ uncertainty 
cited above, $\tau$ lies in the interval [$1.1\times 10^{-9}$~s, $1.7\times 
10^{-8}$~s], exceeding by far the weak-interaction lifetime scale set by the 
free-$\Lambda$ lifetime $\tau_{\Lambda}=2.6\times 10^{-10}$~s, so that the 
robust observation of $\Lambda\Lambda$ hypernuclei by their weak-interaction 
decay modes does not rule out the existence of a deeply bound $H$ dibaryon 
with mass below $m_{\Lambda}+m_n$.

\section{$\Lambda\Lambda$ nonmesonic weak decays} 
\label{sec:nmwd} 

\begin{figure}[!t]
\begin{center}
\includegraphics[width=0.9\textwidth]{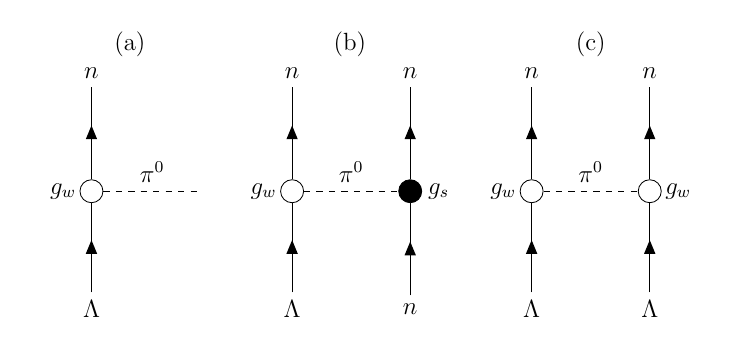}
\caption{$\Delta{\cal S}=1$ $\Lambda\to n$ weak-interaction diagrams in free 
space (a) or in $\Lambda$ hypernuclei (b), see Dalitz~\cite{Dalitz05}, and 
$\Delta{\cal S}=2$ $\Lambda\Lambda\to nn$ weak-interaction diagram (c), all 
involving emission (a) or exchange (b,c) of a $\pi^0$ meson. Weak-interaction 
and strong-interaction coupling constants $g_w$ and $g_s$, respectively, are 
denoted by circles.} 
\label{fig:piwd} 
\end{center} 
\end{figure}

Having realized that a deeply bound $H$ dibaryon lying below $m_{\Lambda}+m_n$ 
is not in conflict with the weak-decay lifetime scale $\tau_{\Lambda}\sim 
10^{-10}$~s of all observed $\Lambda\Lambda$ hypernuclei, we now estimate the 
leading $\Delta{\cal S}=2$ weak-interaction decay rate of $H$, that of the 
$H\to nn$ two-body decay. $H$ is represented here, as above, by its deeply 
bound $\Lambda\Lambda$ component. Although $\Delta{\cal S}=2$ $\Lambda\Lambda
\to nn$ transitions are not constrained directly by experiment, they are 
related to $\Delta{\cal S}=1$ $\Lambda n\to nn$ transitions which are 
constrained by ample lifetime data in $\Lambda$ hypernuclei~\cite{GHM16}. 
These $\Delta{\cal S}=1$ transitions, including the pion-exchange transition 
depicted in Fig.~\ref{fig:piwd}(b), proceed in $\Lambda$ hypernuclei with 
a total rate comparable to the $\Lambda\to n\pi^0$ free-space decay rate 
associated with Fig.~\ref{fig:piwd}(a). The weak-interaction coupling constant 
$g_w$ extracted from the free-space $\Lambda$ lifetime, and proved to be 
relevant in the $\Lambda n\to nn$ nonmesonic decay of $\Lambda$ hypernuclei, 
could then be used as shown in Fig.~\ref{fig:piwd}(c) to estimate the strength 
of the $\Delta{\cal S}=2$ $\Lambda\Lambda\to nn$ weak-decay transition. 

\begin{figure}[!h] 
\begin{center} 
\includegraphics[width=0.5\textwidth]{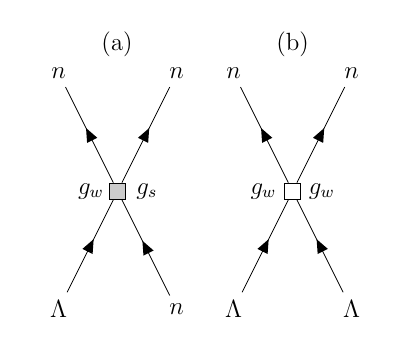} 
\caption{$^1S_0\to {{^1}S_0}$ LO EFT $\Delta{\cal S}\neq 0$ weak-interaction 
diagrams: 
(a) $\Delta{\cal S}=1$ $\Lambda n\to nn$, (b) $\Delta{\cal S}=2$ 
$\Lambda\Lambda\to nn$.} 
\label{fig:nmwd} 
\end{center} 
\end{figure}

Pion exchange is not the only contributor to the nonmesonic weak decay (NMWD) 
of $\Lambda$ hypernuclei. Owing to the large momentum transfer in 
Fig.~\ref{fig:piwd}(b), pion exchange generates mostly a tensor 
$^3S_1\to {{^3}D_1}$ transition which is Pauli forbidden for $nn$, so 
shorter-range meson exchanges need to be considered. However, the next 
candidate of pseudoscalar meson-exchange, $K$ meson-exchange, interferes 
destructively with pion exchange in the $\Lambda n\to nn$ $^1S_0\to {{^1}S_0}$ 
parity-conserving (PC) transition of interest~\cite{PR01}. It is useful then 
to follow an EFT approach initiated by Jun~\cite{Jun01} and applied 
systematically by Parre\~{n}o et al.~\cite{PBH04,Perez11} where 
Figs.~\ref{fig:piwd}(b,c) are supplemented by Figs.~\ref{fig:nmwd}(a,b) 
respectively. The square vertices in these figures stand at leading-order (LO) 
for $^1S_0\to {{^1}S_0}$ low-energy constants (LECs) denoted schematically by 
$g_wg_s$ and $g_w^2$, respectively. These vertices incorporate effects of 
heavier-meson (and thus shorter-range) exchange diagrams which are poorly 
known. Furthermore, we note that the smallness of the $\Lambda$ intrinsic 
asymmetry parameter $a_{\Lambda}$ measured at KEK in the NMWD of \lamb{5}{He} 
and \lamb{12}{C} \cite{KEK07} suggests that the $\Lambda n\to nn$ $^1S_0\to 
{{^3}P_0}$ parity-violating (PV) amplitude, disregarded here, is substantially 
smaller than the $^1S_0\to {{^1}S_0}$ PC amplitude considered below. 

Since the momentum $p_f\approx 420$~MeV/c of each of the final neutrons in 
Fig.~\ref{fig:nmwd}(a) is much larger than the Fermi momentum of the initial 
neutron, the $\Lambda$ hypernuclear decay rate induced by this diagram 
is well approximated by a quasi-free expression tested in studies of $\Sigma$ 
hypernuclear widths~\cite{GD80,Gal82}, 
\begin{equation} 
\Gamma_n = v_{\Lambda n}\,\sigma_{\Lambda n \to nn}\,\frac{1}{4}\rho_n, 
\label{eq:sigma} 
\end{equation} 
where $\rho_n=0.084$~fm$^{-3}$ is the neutron density in nuclear matter and 
$\frac{1}{4}$ stands for the fraction of initial-state neutrons satisfying 
$S_{\Lambda n}=0$. To calculate the $\Delta{\cal S}=1$ two-body reaction cross 
section $\sigma_{\Lambda n\to nn}$ at LO, we use a $\Delta{\cal S}=1$ $^1S_0 
\to {{^1}S_0}$ contact interaction $C_1^{(\lambda)}\delta_{\lambda}(\rvec)$, 
Eq.~(\ref{eq:gaussian}), with a LEC $C_1^{(\lambda)}(\Lambda n)$ determined 
by fitting the r.h.s. of Eq.~(\ref{eq:sigma}) to $\Gamma_n = (0.35\pm 0.04)
\Gamma_{\Lambda}$ where $\Gamma_{\Lambda} = \hbar/\tau_{\Lambda}$. We used 
a value $\Gamma_n/\Gamma_p = 0.55\pm 0.10$ from \lamb{12}{C} NMWD measurements 
(see Table XIII of Ref.~\cite{GHM16}) assuming that all $\Lambda N$ NMWD modes 
sum up to $\Gamma_{\Lambda}$. Note that lifetimes of heavier hypernuclei 
are shorter by $\sim 25\%$ than $\tau_{\Lambda}=263$~ps ($\tau_{hyp}\approx 
210$~ps~\cite{KEK05,JLAB18}) owing most likely to $\Lambda NN$ NMWD 
modes~\cite{GHM16}. 

Evaluating $\sigma_{\Lambda n\to nn}$ at rest, Eq.~(\ref{eq:sigma}) assumes 
the form  
\begin{equation} 
\Gamma_n = \frac{\mu_{nn}\,k_n}{(2\pi\hbar c)^2}\times\frac{\rho_n}{4} \int
{|<\psi_{nn}^{({\kvec}_n)}({\rvec})|C_1^{(\lambda)}\delta_{\lambda}(\rvec)|
\psi_{\Lambda n}(r)>|^2\,{\rm d}{\hat\kvec}_n},  
\label{eq:Gamma_n} 
\end{equation} 
where the initial, at-rest, $\psi_{\Lambda n}$ and final 
$\psi_{nn}^{({\kvec}_n)}$ wavefunctions are given by 
\begin{equation} 
\psi_{\Lambda n}(r)= 1 - \exp(-\frac{1}{6}q_c^2r^2),\,\,\,\,\,
\psi_{nn}^{({\kvec}_n)}({\rvec})=\exp(i{\kvec}_n\cdot{\rvec})\,
(1-j_0(q_c r)),  
\label{eq:psi_nn} 
\end{equation} 
with $k_n$ the neutron momentum release. Note that both repulsive short-range 
initial-state interaction (ISI) $1-\exp(-\frac{1}{6}q_c^2r^2)$ and final-state 
interaction (FSI) $1-j_0(q_c r)$, where $q_c=m_{\omega}=3.97$~fm$^{-1}$, start 
as $\frac{1}{6}q_c^2r^2$ at small r. 

For $\lambda=4$~fm$^{-1}$, Eq.~(\ref{eq:Gamma_n}) yields $C_1^{(\lambda=4)}
(\Lambda n)=-(379\pm 23)$~eV$\cdot$fm$^3$ up to a sign. The assigned 
uncertainty is of a statistical nature owing to that of the underlying value 
$\Gamma_n = (0.35\pm 0.04)\Gamma_{\Lambda}$. As for model dependence, 
switching off either ISI or FSI reduces $C_1$ roughly by a factor of two, 
while switching off both results in $C_{1~{\rm PW}}^{(\lambda=4)}({\Lambda n})
=-(100\pm 6)$~eV$\cdot$fm$^3$, where PW stands for plane waves. A weaker 
dependence on $\lambda$ is found within $\delta\lambda=\pm 1$~fm$^{-1}$ 
about $\lambda=4$~fm$^{-1}$. As for ratios of $\Delta{\cal S}=1$ LECs to 
$\Delta{\cal S}=0$ LECs, using a $\Delta{\cal S}=0$ $^1S_0$ $\Lambda N$ LEC 
$C_0^{(\lambda=4)}(\Lambda N)=-239$ MeV$\cdot$fm$^3$ \cite{CSBGM19} one finds 
$C_{1~{\rm PW}}^{(\lambda=4)}({\Lambda n})/C_0^{(\lambda=4)}(\Lambda N)=
(4.18\pm 0.25)\times 10^{-7}$, larger by less than a factor of two than 
$g_w/g_s=2.21\times10^{-7}$ when $g_w$ is identified with the Fermi 
weak-interaction constant $G_Fm_{\pi}^2=2.21\times 10^{-7}$ and a value 
$g_s=1$, well below the pion-exchange diagram value $g_{\pi NN}\approx 13.6$, 
is adopted. Below we use $G_Fm_{\pi}^2$ to estimate the $\Delta{\cal S}=2$ 
LEC, namely $C_2^{(\lambda)}(\Lambda\Lambda) = 2.21\cdot 10^{-7}\times 
C_1^{(\lambda)}(\Lambda n)$. We assign a factor of two systematical 
uncertainty to this choice by varying $g_s$ between 1/2 and 2 about the 
chosen value $g_s=1$. Altogether, the $H\to nn$ decay rate is given then by 
\begin{equation} 
\Gamma(H\to nn) = 
\frac{\mu_{nn}\,k_n}{(2\pi\hbar c)^2} \int{|<\psi_{nn}^{({\kvec}_n)}({\rvec})|
C_2^{(\lambda)}\delta_{\lambda}(\rvec)|{\tilde{\psi}}_{\Lambda\Lambda}(r)>|^2
\,{\rm d}{\hat\kvec}_n},  
\label{eq:H->nn} 
\end{equation} 
where ${\tilde{\psi}}_{\Lambda\Lambda}(r)$ is defined in Eq.~(\ref{eq:Psi_f}). 

\begin{table}[!h] 
\begin{center} 
\caption{$H\to nn$ decay rate $\Gamma_H$, Eq.~(\ref{eq:H->nn}) with $\lambda = 
4$~fm$^{-1}$, and $H$ lifetime $\tau_H=\hbar/\Gamma_H$ for several choices of 
$H$ binding energy $B_{\Lambda\Lambda}$ and neutron momentum release $k_n$ 
related values. Here, $C_2^{(\lambda)}=(2.21\times 10^{-7})\,C_1^{(\lambda)}$ 
with $C_1^{(\lambda=4)}=-(379\pm 23)$~eV$\cdot$fm$^3$. The listed uncertainty 
follows that of the input value $\Gamma_n = (0.35\pm 0.04)\Gamma_{\Lambda}$, 
see text below Eq.~(\ref{eq:sigma}).} 
\begin{tabular}{cccc} 
\hline 
$B_{\Lambda\Lambda}$ (MeV) & $k_n$ (fm$^{-1}$) & $\Gamma_H$ ($10^{-20}$ eV) & 
$\tau_H$ ($10^5$ s) \\ 
\hline 
176 & 2.109 & 1.57$\pm$0.19 & 0.78$\pm$0.09  \\ 
200 & 1.955 & 1.44$\pm$0.17 & 0.83$\pm$0.10  \\ 
300 & 1.130 & 0.86$\pm$0.10 & 1.35$\pm$0.16  \\ 
\hline 
\end{tabular} 
\label{tab:Hnn} 
\end{center} 
\end{table} 

$H\to nn$ decay rate values $\Gamma_H$ and their associated lifetime values 
$\tau_H$, as calculated using $\lambda=4$~fm$^{-1}$ in Eq.~(\ref{eq:H->nn}), 
are listed in Table~\ref{tab:Hnn}. A weak dependence 
of $\tau_H$ on the $H$ mass ($m_H=2m_{\Lambda}-B_{\Lambda\Lambda}$) is noted, 
in contrast to the strong dependence observed in Table~\ref{tab:I_R} for 
the strong-interaction lifetime of \lamblamb{6}{He} caused by the rapid 
exponential decrease $\exp(-a_{\Phi}^2k_H^2)$ upon increasing $k_H$ in 
Eq.~(\ref{eq:I_R}). The relatively large value $a_{\Phi}\approx 1.5$~fm 
extracted from \lamblamb{6}{He} is replaced here by a considerably smaller 
value of less than 0.3~fm owing to the considerably smaller size parameters 
of the deeply bound $H$ wavefunction $\tilde\psi$ and of the contact term 
$\delta_{\lambda}(\rvec)$, resulting altogether in a weak $k_n$ dependence. 
The calculated $H\to nn$ lifetimes listed in Table~\ref{tab:Hnn} 
are then uniformly of order 10$^5$~s, less than 1~yr, many orders of magnitude 
shorter than cosmological time scales commensurate with the Universe age. 
Regarding the model dependence of the calculated lifetimes, $\tau_H \sim 
10^5$~s, we note that it depends weakly on the range parameter of the LECs 
$C_1^{(\lambda)}$ and $C_2^{(\lambda)}$ within $\delta\lambda=\pm 1$ about 
$\lambda=4$~fm$^{-1}$. Suppressing FSI in $\psi_{nn}^{({\kvec}_n)}({\rvec})$ 
would have increased $\Gamma_H$ by a factor of 5 to 6. However, suppressing 
FSI also (along with ISI) in the extraction of $C_1^{(\lambda)}$ from 
$\Gamma_n$ has an opposite effect; In total, the resulting PW value of 
$\Gamma_H$ is a factor of 2 to 3 {\it smaller} and the PW value of $\tau_H$ 
2 to 3 {\it larger} than listed in Table~\ref{tab:Hnn}. This model dependence 
is of the same scale as the factor of four systematical uncertainty noted 
earlier, arising from the choice of $g_s=1$ in relating $C_2^{(\lambda)}$ to 
$C_1^{(\lambda)}$.

\section{Concluding remarks} 
\label{sec:concl} 

In this work we considered hypernuclear constraints on the existence and 
lifetime of an hypothetical deeply-bound doubly-strange $H$ dibaryon. 
Regarding the existence of $H$, it was found by considering the unambiguously 
identified \lamblamb{6}{He} double-$\Lambda$ hypernucleus~\cite{HN18} that 
its strong-interaction lifetime for decay to $^4{\rm He}+H$ would increase 
substantially upon decreasing $m_H$, exceeding for $m_H < m_{\Lambda}+m_n$ 
by far the 10$^{-10}$~s weak-interaction hypernuclear lifetime scale. Thus, 
the unique observation of double-$\Lambda$ hypernuclei through weak decay to 
single-$\Lambda$ hypernuclei does not rule out on its own the existence of 
a deeply-bound $H$ dibaryon in the mass range $2m_n$ to $m_{\Lambda}+m_n$, 
defying doubts raised by Dalitz et al. 35 years ago~\cite{Dalitz89}, but in 
agreement with the 20 years old claim by FZ~\cite{FZ04}. Special attention was 
given in our evaluation to constrain hadronic cluster sizes by their binding 
energies and, indeed, our own conclusion follows from respecting the large 
difference between the r.m.s. radius of the loosely bound $\Lambda\Lambda$ 
pair in \lamblamb{6}{He} and that of the compact $\Lambda\Lambda$ pair 
within the $H$ dibaryon. And furthermore, the anticipated $\Lambda\Lambda$ 
short-range repulsion was fully considered by incorporating the SRC factor 
$(1-j_0(\kappa r))$, Eq.~(\ref{eq:phiLL}), into the $\Lambda\Lambda$ 
wavefunction. 
  
Provided an $H$-like dibaryon exists in the mass range $2m_n$ to $m_{\Lambda} 
+m_n$, it was found that its $\Delta{\cal S}=2$ decay lifetime $\tau(H\to nn)$ 
would be quite long, of the order of $10^5$~s, but many orders of magnitude 
shorter than cosmological lifetimes comparable to the age of the Universe that 
$H$ would need to qualify for a DM candidate. To reach this conclusion we used 
a weak-interaction EFT approach~\cite{PBH04} constrained by experimentally 
known, largely nonmesonic, hypernuclear lifetimes~\cite{GHM16}. Our conclusion 
is in stark disagreement with that reached by FZ~\cite{FZ04} using outdated by 
now hard-core strong-interaction nuclear models and, furthermore, disregarding 
the constraint imposed on the $H$ radius $r_H$ by its binding energy. 

We have not considered in the present work the case for an $H$-like dibaryon 
with mass below the $nn$ threshold, a scenario likely to be ruled out 
by the established stability of several key nuclei, notably $^{16}$O. 
A straightforward calculation of the hypothetical two-body decay rate 
$\Gamma({^{16}{\rm O}}\to H + {^{14}{\rm O}})$ gives indeed a rate many 
orders of magnitude larger than the upper bound established for oxygen by 
Super-Kamiokande~\cite{SK15}. A report of this calculation is in preparation.

\section*{Acknowledgements}
Special thanks are due to Nir Barnea and Martin Sch{\"a}fer for useful 
discussions during early stages of this work, and to Martin Sch{\"a}fer 
for providing me with unpublished numerical output extracted from 
Ref.~\cite{CSBGM19}. This work is part of a project funded by the EU 
Horizon 2020 Research \& Innovation Programme under grant agreement 824093.


\begin{thebibliography}{99}

\bibitem{AA20} N.~Aghanim, et al. (Planck Collab.), 
\emph{Planck 2018 results}, Astronomy \& Astrophysics 641 (2020) A6. 

\bibitem{LQCD18} S.~Gongyo, et al. (HAL QCD Collab.), \emph{Most Strange 
Dibaryon from Lattice QCD}, Phys. Rev. Lett. 120 (2018) 212001. 

\bibitem{LQCD11a} S.R.~Beane, et al. (NPLQCD Collab.), \emph{Evidence for a 
Bound $H$ Dibaryon from Lattice QCD}, Phys. Rev. Lett. 106 (2011) 162001. 

\bibitem{LQCD11b} T.~Inoue, et al. (HAL QCD Collab.), \emph{Bound $H$ Dibaryon 
in Flavor SU(3) Limit of Lattice QCD}, Phys. Rev. Lett. 106 (2011) 162002. 

\bibitem{LQCD21} J.R.~Green, A.D.~Hanlon, P.M.~Junnarkar, H.~Wittig, 
\emph{Weakly Bound $H$ Dibaryon from SU(3)-Flavor-Symmetric QCD}, Phys. Rev. 
Lett. 127 (2021) 242003, and references listed therein to earlier LQCD 
searches for a bound $H$ dibaryon. 

\bibitem{LQCD11c} P.E.~Shanahan, A.W.~Thomas, R.D.~Young, \emph{Mass of the 
$H$ Dibaryon}, Phys. Rev. Lett. 107 (2011) 092004. 

\bibitem{LQCD12} T.~Inoue, et al. (HAL QCD Collab.), \emph{Two-baryon 
potentials and $H$-dibaryon from 3-flavor lattice QCD simulations}, 
Nucl. Phys. A 881 (2012) 28. 

\bibitem{EFT12} J.~Haidenbauer, U.-G.~Mei{\ss}ner, \emph{Exotic bound states 
of two baryons in light of chiral effective field theory}, Nucl. Phys. A 881 
(2012) 44. 








\bibitem{Jaffe77} R.L.~Jaffe, \emph{Perhaps a stable Dihyperon}, 
Phys. Rev. Lett. 38 (1977) 195. 

\bibitem{AGS78} A.S.~Carroll, et al., \emph{Search for Six-Quark States}, 
Phys. Rev. Lett. 41 (1978) 777. 

\bibitem{Belle13} B.H.~Kim, et al. (Belle Collaboration), \emph{Search 
for an H Dibaryon with a Mass near $2m_{\Lambda}$ in $\Upsilon(1S)$ and 
$\Upsilon(2S)$ Decays}, Phys. Rev. Lett. 110 (2013) 222002.  

\bibitem{ALICE16} J.~Adam, et al. (ALICE Collaboration), \emph{Search for 
weakly decaying $\bar{~\Lambda n~}$ and $\Lambda\Lambda$ exotic bound states 
in central Pb-Pb collisions at $\sqrt{s_{NN}}$=2.76 TeV}. Phys. Lett. B 752 
(2016) 267. 

\bibitem{BABAR19} J.P.~Lees, et al. ($BaBar$ Collaboration), \emph{Search 
for a Stable Six-Quark State at BABAR}, Phys. Rev. Lett. 122 (2019) 072002. 

\bibitem{Dalitz89} R.H.~Dalitz, D.H.~Davis, P.H.~Fowler, A.~Montwill, 
J.~Pniewski, J.A.~Zakrzewski, \emph{The identified $\Lambda\Lambda$ 
hypernuclei and the predicted H-particle}, Proc. R. Soc. Lond. A 426 (1989) 1. 

\bibitem{GHM16} A.~Gal, E.V.~Hungerford, D.J.~Millener, \emph{Strangeness in 
nuclear physics}, Rev. Mod. Phys. 88 (2016) 035004. 

\bibitem{HN18} E.~Hiyama, K.~Nakazawa, Annu. Rev. Nucl. Part. Sci. 68 (2018) 
131. 

\bibitem{Ahn13} J.K.~Ahn, et al. (E373 KEK-PS Collaboration), 
\emph{Double-$\Lambda$ hypernuclei observed in a hybrid emulsion experiment}, 
Phys. Rev. C 88 (2013) 014003. 

\bibitem{Gal13} A.~Gal, \emph{Comment on ``Strangeness $-$2 Hypertriton"}, 
Phys. Rev. Lett. 110 (2013) 179201. 

\bibitem{Farrar03} G.R.~Farrar, \emph{A Stable H-Dibaryon: Dark Matter 
Candidate Within QCD?}, Int'l J. Theor. Phys. 42 (2003) 1211. 

\bibitem{FW23} G.R.~Farrar, Z.~Wang, \emph{Constraints on long-lived 
di-baryons and di-baryonic dark matter}, arXiv:2306.03123 [hep-ph] and 
references to earlier work cited therein.

\bibitem{SK15} J.~Gustafson, et al. (Super-Kamiokande Collaboration), 
\emph{Search for dinucleon decay into pions at Super-Kamiokande}, 
Phys. Rev. D 91 (2015) 072009. 




\bibitem{DGH86} J.F.~Donoghue, E.~Golowich, B.R.~Holstein, \emph{Weak decays 
of the $H$ dibaryon}, Phys. Rev. D 34 (1986) 3434. 

\bibitem{FZ04} G.R.~Farrar, G.~Zaharijas, \emph{Nuclear and nucleon 
transitions of the H dibaryon}, Phys. Rev. D 70 (2004) 014008.  


\bibitem{CSBGM19} L.~Contessi, M.~Sch\"{a}fer, N.~Barnea, A.~Gal, 
J.~Mare\v{s}, \emph{The onset of $\Lambda\Lambda$ hypernuclear binding}, 
Phys. Lett. B 797 (2019) 134893. 

\bibitem{CBG18} L.~Contessi, N.~Barnea, A.~Gal, \emph{Resolving the 
$\Lambda$ Hypernuclear Overbinding Problem in Pionless Effective Field 
Theory}, Phys. Rev. Lett. 121 (2018) 102502. 



\bibitem{Inoue19} T.~Inoue (for the HAL QCD Collaboration), \emph{Strange 
nuclear physics from OCD on lattice}, AIP Conf. Proc. 2130 (2019) 020002. 

\bibitem{MPR18} J.~Maneu, A.~Parre{\~n}o, A.~Ramos, \emph{Effects of 
$\Lambda\Lambda-\Xi N$ mixing in the decay of ${\cal S}=-2$ hypernuclei}, 
Phys. Rev. C 98 (2018) 025208. 



\bibitem{KW24} N.~Kaiser, W.~Weise, \emph{Sizes of the Nucleon}, Phys. Rev. C 
110 (2024) 015202. 

\bibitem{Baym69} G.~Baym, \emph{Lectures on Quantum Mechanics}, W.A.~Benjamin 
1969 (Library of Congress Catalog Card 68-5611), in particular Eq.~(12-33) and 
related text. 

\bibitem{Sasaki20} K.~Sasaki, et al. (HAL QCD Collab.), \emph{$\Lambda\Lambda$ 
and $N\Xi$ interactions from lattice QCD near the physical point}, 
Nucl. Phys. A 998 (2020) 121737. 



\bibitem{Dalitz05} R.H.~Dalitz, \emph{50 years of hypernuclear physics}, 
Nucl. Phys. A 754 (2005) 14c. 

\bibitem{PR01} A.~Parre\~{n}o, A.~Ramos, \emph{Final-state interactions in 
hypernuclear decay}, Phys. Rev. C 65 (2001) 015204.

\bibitem{Jun01} J.-H.~Jun, \emph{Four-baryon point $\Lambda N \rightarrow NN$ 
interaction for the nonmesonic weak decay of the hypernuclei \lamb{4}{H}, 
\lamb{4}{He}, \lamb{5}{He}, and \lamb{12}{C}}, Phys. Rev. C 63 (2001) 044012. 
An earlier report was given at HYP1997, Nucl. Phys. A 639 (1997) 337c. 

\bibitem{PBH04} A.~Parre\~{n}o, C.~Bennhold, B.R.~Holstein, \emph{$\Lambda N 
\to NN$ weak interaction in effective-field theory}, Phys. Rev. C 70 (2004) 
051601(R); \emph{An EFT for the weak $\Lambda N$ interaction}, Nucl. Phys. A 
754 (2005) 127c. 

\bibitem{Perez11} A.~P\'{e}rez-Obiol, A.~Parre\~{n}o, B.~Juli\'{a}-D\'{i}az, 
\emph{Constraints on effective field theory parameters for the $\Lambda N\to 
NN$ transition}, Phys. Rev. C 84 (2011) 024606. 

\bibitem{KEK07} T.~Maruta, et al., \emph{Decay asymmetry in NMWD of light 
$\Lambda$-hypernuclei}, Eur. Phys. J. A 33 (2007) 255. 

\bibitem{GD80} A.~Gal, C.B.~Dover, \emph{Narrow $\Sigma$-Hypernuclear States}, 
Phys. Rev. Lett. 44 (1980) 379. 

\bibitem{Gal82} A.~Gal, \emph{Are $\Sigma$ Nuclear States Really Narrow?}, 
Proc. Int'l. Conf. on Hypernuclear and Kaon Physics, Heidelberg, Germany, 
1982, ed. B.~Povh (MPI H - 1982 - V 20) pp. 27-36.

\bibitem{KEK05} Y.~Sato, et al. (KEK-PS E307 Collaboration), \emph{Mesonic 
and nonmesonic weak decay widths of medium-heavy $\Lambda$ hypernuclei}, 
Phys. Rev. C 71 (2005) 025203. 

\bibitem{JLAB18} X.~Qiu, L.~Tang, et al. (HKS JLab E02-017 Collaboration), 
\emph{Lifetime of heavy hypernuclei}, Nucl. Phys. A 973 (2018) 116.  









\end{thebibliography}
\end{document}